\documentclass[journal]{IEEEtran}

\usepackage{amsmath}

\usepackage{algpseudocode} 
\usepackage{algorithm}     

\ifCLASSINFOpdf
\usepackage{graphicx}
\graphicspath{{../images/}}
\DeclareGraphicsExtensions{.pdf,.jpeg,.png}
\else
\usepackage[dvips]{graphicx}
\graphicspath{{../images/}}
\DeclareGraphicsExtensions{.eps}
\fi

\usepackage{cite}

\usepackage{url}
\usepackage{orcidlink}

\usepackage{makecell}
\usepackage{booktabs}
\usepackage{orcidlink}
\usepackage{diagbox}

\begin{document}
	%
	\title{Grid-Agent: An LLM-Powered Multi-Agent System for Power Grid Control}
	%
	%
	
	\author{
		Yan~Zhang\orcidlink{0000-0001-9867-4924},
		Ahmad~Mohammad~Saber\orcidlink{0000-0003-3115-2384},~\IEEEmembership{Member,~IEEE,}
		Amr~Youssef\orcidlink{0000-0002-4284-8646},~\IEEEmembership{Senior Member,~IEEE,}
		and~Deepa~Kundur\orcidlink{0000-0001-5999-1847},~\IEEEmembership{Fellow,~IEEE}
		\thanks{Yan~Zhang, Ahmad Mohammad Saber, and Deepa Kundur are with 
			the Department of Electrical and Computer Engineering, University of Toronto, Toronto, ON M5S 1A1, Canada
			(e-mails: 
			\href{mailto:claudeyan.zhang@mail.utoronto.ca}{claudeyan.zhang@mail.utoronto.ca};
			\href{mailto:ahmad.m.saber@ieee.org}{ahmad.m.saber@ieee.org};
			\href{mailto:dkundur@ece.utoronto.ca}{dkundur@ece.utoronto.ca}).}
		\thanks{Amr Youssef is with
			the Concordia Institute for Information Systems Engineering,
			(CIISE), 
			Concordia University,
			Montreal, QC H3G 1M8, 
			Canada
			(e-mail: 
			\href{mailto:youssef@ciise.concordia.ca}{youssef@ciise.concordia.ca}).}
	}

	\markboth{Preprint. Under review. Do not distribute.}%
	{Authors \MakeLowercase{\textit{et al.}}: Semantic Reasoning Meets Numerical Precision: An LLM-Powered Multi-Agent System for Power Grid Control}
	
	\maketitle
	
	\begin{abstract}
		
		Modern power grids face unprecedented complexity from Distributed Energy Resources (DERs), Electric Vehicles (EVs), and extreme weather, while also being increasingly exposed to cyberattacks that can trigger grid violations. 
		This paper introduces Grid-Agent, an autonomous AI-driven framework that leverages Large Language Models (LLMs) within a multi-agent system to detect and remediate violations. Grid-Agent integrates semantic reasoning with numerical precision through modular agents: a planning agent generates coordinated action sequences using power flow solvers, while a validation agent ensures stability and safety through sandboxed execution with rollback mechanisms. To enhance scalability, the framework employs an adaptive multi-scale network representation that dynamically adjusts encoding schemes based on system size and complexity. Violation resolution is achieved through optimizing switch configurations, battery deployment, and load curtailment. Our experiments on IEEE and CIGRE benchmark networks, including the IEEE 69-bus, CIGRE MV, IEEE 30-bus test systems, demonstrate superior mitigation performance, highlighting Grid-Agent’s suitability for modern smart grids requiring rapid, adaptive response.

	\end{abstract}
	
	
	\IEEEpeerreviewmaketitle
	
	\section{Introduction}
	\label{sec:introduction}
	\IEEEPARstart{M}{odern} power grids face unprecedented challenges due to increased integration of renewable energy, electric vehicles (EVs), increasing frequency of extreme weather events, and aging infrastructure. These factors demand optimization strategies that are not only intelligent and responsive, but also adaptable to highly dynamic and uncertain operating conditions, especially as grid failures and blackouts have been estimated to cost the U.S. economy over \$150 billion annually \cite{power_outage_cost}. Traditional power system optimization methods are based on fixed rule-based logic and deterministic numerical algorithms, often fall short in their ability to generalize, interpret contextual information, or reason semantically—capabilities that are essential for managing the complexity of real-world grid scenarios.
	
	The emergence of Large Language Models (LLMs) presents transformative opportunities for power system applications by enabling the integration of natural language reasoning with numerical precision. Recent advancements in LLM-based multi-agent systems have demonstrated impressive capabilities in solving complex, multi-faceted problems across various domains \cite{multi_agent_llm_2024} \cite{multi_agent_for_problem_solving}. However, their application to mission-critical infrastructures such as electric power grids remains largely unexplored.
	
	This paper introduces Grid-Agent, an autonomous, AI-driven framework conceived as an exploration into how LLMs can be applied to power system control, rather than as a finalized, production-ready solution. The primary goal is to push the boundaries of what LLMs and multi-agent systems might bring to future smart grid operations by demonstrating a novel methodology for autonomous grid management. Consequently, this work prioritizes the exploration of agentic reasoning, coordinated planning, and semantic interpretation over the formal optimality of the resulting control actions. The focus is on demonstrating a paradigm where Grid-Agent can interpret network topology, detect electrical violations, and formulate remediation strategies by integrating semantic understanding with domain-specific analysis tools.
	The key contribution of this work includes:
	
	\begin{itemize}
		\item Introduction of Grid-Agent, a modular multi-agent framework that combines LLM-based semantic reasoning with numerical solvers to autonomously detect and remediate grid violations. To ensure scalability across diverse network sizes, Grid-Agent incorporates an adaptive multi-scale representation that dynamically adjusts encoding granularity. Violation resolution is formulated as a coordinated optimization problem that integrates topology reconfiguration, battery scheduling, and demand response to maximize effectiveness while minimizing the number of control actions. Furthermore, the framework embeds a continuous learning pipeline that generates training data and fine-tunes its LLM components, enabling performance improvement and adaptation to evolving grid conditions over time.
		\item To validate the above contribution, Grid-Agent was implemented and tested on benchmark IEEE and CIGRE systems, including the IEEE 69-bus, the IEEE 30-bus, and the CIGRE MV test systems. Our experimental results demonstrate its superior ability to mitigate violations compared with baseline approaches, its scalability across networks of different sizes and complexities, and its robustness against multiple simultaneous violations, including those arising from cyberattack-induced contingencies.
	\end{itemize}
	
	
	The remainder of this paper is organized as follows: Section \ref{sec:related_work} reviews related work in power systems optimization and LLM-based multi-agent systems. Section \ref{sec:preliminaries} formulates the power grid violation resolution problem. Section \ref{sec:system_design} presents the Grid-Agent system architecture and methodology. Section \ref{sec:setup} describes the experimental setup and validation. Section \ref{sec:discussion} discusses results and future works. Section \ref{sec:conclusion} concludes the achievements.
	
	\section{Related Work}
	\label{sec:related_work}
	
	\subsection{Power Systems Optimization}
	
	Traditional power system optimization has progressed from manual control to sophisticated automated frameworks. Classical methods, such as Optimal Power Flow (OPF) \cite{opf_steady_state}, \cite{opf_survey}, have been widely applied to objectives like peak shaving, cost/loss minimization, and violation mitigation. While mathematically robust, these approaches often face convergence issues due to the complexity of modern grids, which incorporate high levels of distributed renewable energy, highly variable loads, and numerous discrete controllable assets such as switches and tap changers. These factors result in a mixed-integer, non-convex, nonlinear, and unbalanced AC optimal powerflow problem, which is NP-hard and computationally intensive to solve. Furthermore, their rigid mathematical structure makes it difficult to incorporate heuristic operator knowledge or adapt to unforeseen event types. Furthermore, these approaches often address isolated tasks and lack a unified framework for holistic, system-level optimization and control. This highlights the need for a system that can integrate high-level strategic reasoning with precise numerical validation, a core principle of our proposed Grid-Agent framework.
	Recent advances in artificial intelligence (AI) have brought machine learning techniques into power systems, with promising results in tasks such as load forecasting \cite{ml_load_forecast}, renewable generation prediction \cite{renewable_prediction}, and fault detection \cite{ai_fault_detection}. However, these approaches generally address isolated tasks and lack a unified framework for holistic, system-level optimization and control.

	\subsection{Large Language Models in Engineering Applications}
	
	The application of LLMs to engineering domains has gained significant attention following the success of models like GPT-4 and Claude. Recent work has explored LLM applications in software engineering \cite{llm_software}, materials science \cite{llm_materials}, and system design \cite{llm_system_design}.
	In the energy sector, preliminary investigations have examined LLMs for grid analytics \cite{llm_grid_analytics}, energy trading \cite{llm_energy_trading}, and demand response optimization \cite{llm_demand_response}. However, these applications primarily focus on data analysis and prediction rather than direct system control and optimization. This leaves a critical gap in utilizing LLM reasoning for active, real-time control and optimization in critical infrastructure, which is the primary focus of our work.
	
	\subsection{Multi-Agent Systems with LLMs}
	
	The integration of LLMs with multi-agent systems is an emerging and rapidly advancing area of research. Recent studies have demonstrated the potential of LLM-based multi-agent frameworks in domains such as software engineering \cite{multi_agent_software}, automated scientific research \cite{multi_agent_research}, and collaborative problem-solving \cite{multi_agent_collaboration}. These systems leverage the reasoning capabilities of LLMs while maintaining structured workflows and state management.
	However, the application of LLM-based multi-agent systems to critical infrastructure domains like power grids presents unique challenges including safety requirements, real-time constraints, and the need for explainable decisions in mission-critical environments.
	
	\section{Grid-Agent: Preliminaries}
	\label{sec:preliminaries}
	
	This section formally defines the power grid violation resolution problem, the underlying physical laws governing the network, and the discrete and continuous control actions available to the agent.
	
	\subsection{Power Grid Violation Resolution Problem}
	We model the power network as a graph ${G} = ({N}, {E})$, where ${N} = \{1, 2, \dots, n\}$ is the set of buses (nodes) and ${E}$ is the set of transmission lines and transformers (edges). The state of the grid is defined by the voltage magnitude $V_i$ and angle $\delta_i$ at each bus $i \in {N}$, along with the active power $P_{ij}$, reactive power $Q_{ij}$, and current flow $I_{ij}$ for each line $(i, j) \in {E}$.
	A \textit{violation} $\mathcal{V}$ occurs when the system state breaches
	predefined operational limits. We consider three primary categories of
	violations: voltage violations, thermal violations, and disconnected buses.
	A \textit{voltage violation} $\mathcal{V}_{\text{volt}}$ occurs when a bus
	voltage magnitude $V_i$ falls outside its acceptable range
	$[V_{\min}, V_{\max}]$. Formally, the set of buses experiencing voltage
	violations is defined as
	
	\begin{equation}
		\mathcal{V}_{\text{volt}} = \{ i \in N \mid V_i < V_{\min}
		\text{ or } V_i > V_{\max} \}
	\end{equation}
	
	\noindent Moreover, a \textit{thermal violation} $\mathcal{V}_{\text{therm}}$ arises when the
	apparent power flow $S_{ij}$ or current $I_{ij}$ on a line exceeds its maximum
	rating. This can be expressed as
	
	\begin{equation}
		\mathcal{V}_{\text{therm}} = \{ (i, j) \in E \mid
		S_{ij} > S_{ij}^{\max} \text{ or } I_{ij} > I_{ij}^{\max} \}
	\end{equation}
	
	\noindent Finally, a \textit{disconnected bus violation} $\mathcal{V}_{\text{disc}}$
	occurs when a bus becomes electrically isolated from the main grid, often
	resulting in a complete loss of power supply.
	
	\subsection{Power Flow Constraints}
	Any valid solution must satisfy the fundamental power flow equations, which represent the physical laws of electricity. For each bus $i \in {N}$, the net active power ($P_i$) and reactive power ($Q_i$) injections must be balanced. These nonlinear equations connect the bus voltages and angles across the entire network:
	
	\begin{equation}
		P_i = V_i \sum_{j \in {N}} V_j [G_{ij} \cos(\delta_i - \delta_j) + B_{ij} \sin(\delta_i - \delta_j)]
	\end{equation}
	
	\begin{equation}
		Q_i = V_i \sum_{j \in {N}} V_j [G_{ij} \sin(\delta_i - \delta_j) - B_{ij} \cos(\delta_i - \delta_j)]
	\end{equation}
	
	\noindent where $G_{ij}$ and $B_{ij}$ are the conductance and susceptance terms from the network's bus admittance matrix, respectively. These equations ensure that any proposed set of control actions results in a physically possible grid state.
	
	\subsection{Action Space Constraints}
	The control actions available to Grid-Agent are constrained by the physical and
	operational capabilities of the grid equipment. The action space consists of
	switch operations, load curtailment, and battery placement with dispatch.
	Switch operations are represented as binary decisions to open or close a
	switchable line $(i,j)$ within the set of designated switchable lines
	${E}_{\text{sw}} \subseteq {E}$. For each switchable line, the state variable is
	defined as
	
	\begin{equation}
		s_{ij} \in \{0, 1\}, \quad \forall (i, j) \in {E}_{\text{sw}}.
	\end{equation}
	
	\noindent Herein, load curtailment is modeled as a continuous reduction in the active power demand
	$P_{i}^{\text{load}}$ at buses belonging to the set of curtailable loads
	${N}_{\text{L}} \subseteq {N}$. The reduction is parameterized by a curtailment
	fraction $\gamma_i$, resulting in a new load value given by
	
	\begin{equation}
		P_{i}^{\text{new\_load}} = P_{i}^{\text{orig\_load}} \cdot (1 - \gamma_i)
	\end{equation}
	
	\begin{equation}
		0 \le \gamma_i \le \gamma_{\max}, \quad \forall i \in N_{\text{L}}
	\end{equation}

	\noindent Further, battery placement and dispatch involve two coupled decisions: a discrete choice
	of where to place batteries, and a continuous choice of how much active and
	reactive power to inject or absorb. Placement is modeled as a binary decision
	variable $b_i$ for each bus, constrained by the total number of available
	batteries $B_{\text{total}}$:
	
	\begin{equation}
		\sum_{i \in {N}} b_i \le B_{\text{total}}, \quad \text{with } b_i \in \{0, 1\}.
	\end{equation}
	
	\noindent Once placed, each battery can dispatch active power $P_{B,i}$ and reactive power
	$Q_{B,i}$, subject to its maximum power rating $S_{B}^{\max}$. Importantly,
	dispatch is only allowed if a battery is installed at the corresponding bus
	($b_i = 1$). The coupling between placement and operation is enforced as
	\begin{equation}
		P_{B,i}^2 + Q_{B,i}^2 \le (b_i \cdot S_{B}^{\max})^2
	\end{equation}
	
	\begin{equation}
		-P_{B}^{\max} \le P_{B,i} \le P_{B}^{\max}
	\end{equation}
	
	\begin{equation}
		-Q_{B}^{\max} \le Q_{B,i} \le Q_{B}^{\max}
	\end{equation}
	
	\noindent The above formulation guarantees that batteries are both allocated and dispatched
	within their physical limits, ensuring they can effectively support grid
	operations.

	\subsection{Large Language Models and Agentic AI}
	
	Solving the violation resolution problem requires more than just numerical
	optimization; it demands a system that can reason about complex, multi-faceted
	scenarios, interpret heterogeneous data, and formulate strategic plans. This is
	where Large Language Models (LLMs) and agentic AI frameworks provide a
	transformative approach.
	
	\subsubsection{LLMs as Reasoning Engines for Grid Management}
	Modern LLMs function as powerful \textit{reasoning engines}. In the context of
	power systems, they offer several key advantages over traditional algorithms.
	First, LLMs bring semantic understanding, as they can process and interpret not
	just raw numerical data, but also heterogeneous sources of information such as
	network topology descriptions, operational logs, and maintenance reports. This
	capability enables a more holistic understanding of grid states. Second, they
	enable contextual reasoning by incorporating high-level domain knowledge and
	operational constraints, such as prioritizing critical loads or avoiding
	switching near vulnerable points, which are difficult to encode in conventional
	optimization models. Finally, recent advancements allow LLMs to operate in a
	tool-augmented fashion, delegating complex numerical calculations to validated
	power flow solvers. This integration grounds their high-level reasoning in the
	physical realities and precise operational constraints of the grid.
	
	\subsubsection{Multi-Agent Systems for Decomposed Problem Solving}
	A single monolithic system struggles to manage the entire scope of violation
	resolution. \textit{Agentic AI} addresses this limitation by decomposing the
	problem into specialized sub-tasks, each handled by a dedicated agent. This
	approach provides several advantages. Different agents can be designed to
	develop specialized expertise, excelling at specific roles such as topology
	analysis, action planning, solution validation, or results summarization. In
	addition, agents can collaborate iteratively, with a Planner agent proposing an
	action and a Validator agent simulating its impact to provide feedback. This
	iterative refinement process allows strategies to improve progressively. Finally,
	the modular structure offers scalability and interpretability: it adapts to
	grids of varying complexity while providing transparent and auditable reasoning
	chains through explicit agent interactions.
	
	Based on the above, the multi-agent system can be formulated as
	
	\begin{equation}
		W = (A, F, C),
	\end{equation}
	
	\noindent where $W$ represents the workflow, $A = \{a_1, a_2, \ldots, a_m\}$ is the set of
	agents, $F$ defines the flow relationships between agents, and $C$ specifies
	conditional branching logic. The workflow state $S_t$ at iteration $t$ captures
	the complete system state, including network configuration, detected violations,
	executed actions, and intermediate results. State transitions are governed by
	
	\begin{equation}
		S_{t+1} = \mathcal{T}(S_t, a_i, \theta),
	\end{equation}
	
	\noindent where $\mathcal{T}$ is the transition function, $a_i$ is the executing agent,
	and $\theta$ represents the agent's parameters and decision logic. This
	formulation enables systematic exploration of the action space while maintaining
	solution quality and computational efficiency through intelligent agent
	coordination.
	
	\section{Grid-Agent: Proposed System Architecture and Methodology}
	\label{sec:system_design}
	
	The Grid-Agent framework is an autonomous multi-agent system designed to resolve power grid violations by integrating the semantic reasoning of Large Language Models (LLMs) with the numerical precision of validated power system simulations. The methodology is founded on a state-driven workflow where specialized agents collaborate to plan, execute, and validate control actions in a secure, sandboxed environment. This section details the four core pillars of this methodology: the multi-agent workflow, the LLM-guided action planning process, the adaptive network representation scheme, and the integrated safety mechanisms.
	
	\subsection{State-Driven Multi-Agent Workflow}
\begin{figure*}[t!]
		\centering
		\includegraphics[width=1.8\columnwidth]{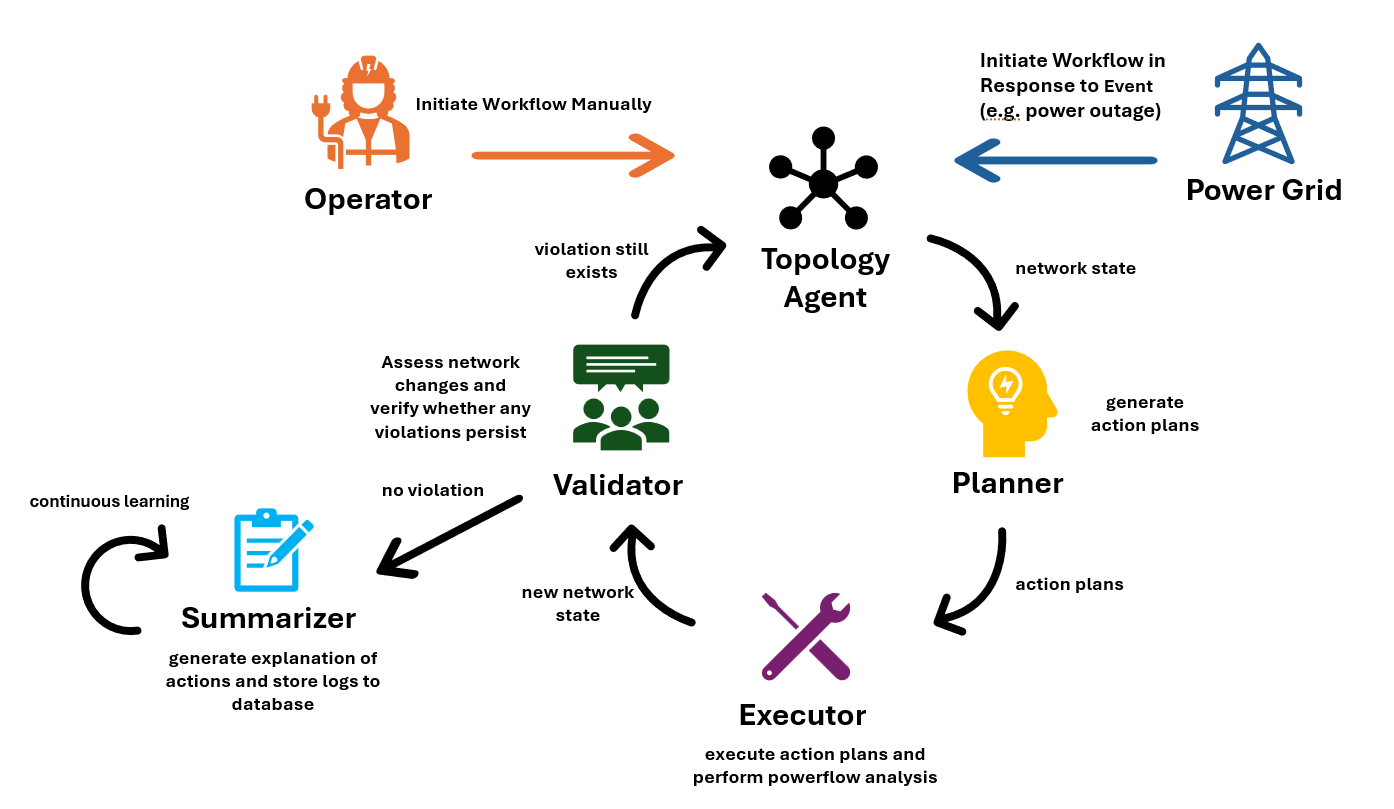}
		\caption{Conceptual diagram of the Proposed Grid Agent}
		\label{fig:architecture}
	\end{figure*}

	The framework's operation is governed by a cyclic, state-driven workflow, depicted in Fig. \ref{fig:architecture}, which is orchestrated by five specialized agents. The process begins with the \textit{Topology Agent}, which initializes the resolution process by parsing the grid's topology and operational data to establish the initial network state and identify existing violations. This foundational context is then passed to the \textit{Planner Agent}, the core reasoning engine, which leverages its LLM-based intelligence to formulate a strategic, multi-step action plan designed to cohesively resolve the identified issues. Once a plan is formulated, the \textit{Executor Agent} acts as the interface to the numerical simulation, translating the Planner's abstract directives into concrete API calls for a power flow solver on a sandboxed copy of the network. Following execution, the \textit{Validator Agent} ensures the efficacy and safety of the plan by running a new analysis to assess whether violations were resolved without introducing new instabilities. Based on this outcome, it either approves the changes and terminates the workflow or reverts the changes and signals the need for a new planning cycle, ensuring monotonic progress. Finally, upon successful resolution, the \textit{Summarizer Agent} facilitates explainability by generating a human-readable summary of the solution and structures the entire process into a data entry for continuous learning datasets.
	
	\subsection{LLM-Guided Coordinated Action Planning}
	The central task of finding an optimal action sequence is navigated by the Planner Agent. Unlike traditional optimizers, Grid-Agent uses the LLM as a tool-augmented reasoning engine. This is achieved through a comprehensive system prompt that structures the LLM's task, defining five key elements:
	The core of this process is the \texttt{Planner Agent}, which formulates a multi-step plan (Algorithm~\ref{alg:llm_agent}).The agent's reasoning is guided by a structured system prompt that specifies:
	1) \textit{Role Definition:} The LLM is instructed to act as an expert power system operator, grounding its responses in domain-specific best practices.
	2) \textit{State Context:}  It receives a structured representation of the current network state, including topology, component statuses, and a detailed list of all active violations.
	3) \textit{Action Space and Constraints:} The prompt explicitly defines the set of available actions (e.g., \texttt{update\_switch\_status}, \texttt{add\_battery}, \texttt{curtail\_load}) and their associated operational constraints (e.g., maximum number of deployable batteries, available curtailable loads).
	4) \textit{Strategic Guidance:} The LLM is guided by a prioritized action policy that mirrors industry best practices: (1) Topology Reconfiguration, (2) Battery Deployment/Dispatch, and (3) Load Curtailment. This constrains the search space and aligns the LLM's reasoning with established operational logic.
	5) \textit{Output Schema:} To ensure reliable machine integration, the LLM is required to respond exclusively with a machine-parsable list of tool calls in a structured format (e.g., JSON). This eliminates ambiguous natural language and enables direct interfacing with the \texttt{Executor Agent}.

	\begin{algorithm}[t!]
		\caption{LLM-Based Multi-Agent Power Grid Optimization}
		\label{alg:llm_agent}
		\begin{algorithmic}[1]
			\Require Network $N$, Maximum iterations $T_{max}$, Available actions $\mathcal{A}_{available}$
			\Ensure Optimized network state $N^*$
			
			\State $N_{work} \leftarrow$ Initialize sandboxed environment($N$)
			\State $\mathcal{V} \leftarrow$ Analyze violations($N_{work}$)
			\State $iter \leftarrow 0$
			
			\While{$\neg IsResolved(\mathcal{V})$ \textbf{and} $iter < T_{max}$}
			
			\State // Multi-Agent Planning Phase
			\State $\mathcal{C} \leftarrow$ Generate Network representation($N_{work}$, $\mathcal{V}$)
			\State $\mathcal{A} \leftarrow$ LLM planning($\mathcal{C}$, $\mathcal{A}_{available}$)
			
			\State // Action Execution Phase
			\For{$action \in \mathcal{A}$}
			\State Execute action($action$, $N_{work}$)
			\State $\mathcal{V}_{new} \leftarrow$ Analyze violations($N_{work}$)
			\If{$IsResolved(\mathcal{V}_{new})$}
			\State \textbf{break}
			\EndIf
			\EndFor
			
			\State // Learning and Adaptation
			\State $effectiveness \leftarrow$ Evaluate actions($\mathcal{V}$, $\mathcal{V}_{new}$, $\mathcal{A}$)
			\State Update Knowledge($effectiveness$)
			
			\State $\mathcal{V} \leftarrow \mathcal{V}_{new}$
			\State $iter \leftarrow iter + 1$
			\EndWhile
			
			\State $explanation \leftarrow$ Generate explanation($N_{work}$, $\mathcal{A}_{all}$)
			\State \Return $N_{work}$, $explanation$
			
		\end{algorithmic}
	\end{algorithm}

	\subsection{Adaptive Multi-Scale Network Representation}
	A primary challenge in applying LLMs to large-scale networks is the constraint of the model's context window. To overcome this, Grid-Agent employs an adaptive network representation scheme that dynamically balances information fidelity with computational tractability. The framework automatically selects the level of detail based on network size and complexity.
	This is performed on two scales: 1) \textit{Full-Component Detail} and  2) \textit{Semantic Graph Abstraction}.
	On the full-component detail scale, for smaller networks, the system provides a complete serialization of every component (buses, lines, loads, generators), offering the LLM comprehensive data for fine-grained analysis.
	On the semantic-graph abstraction scale, for larger networks or scenarios with clustered violations, the system generates a semantic graph representation. This abstraction summarizes healthy sections of the grid and focuses on the electrical relationships between violated components and nearby controllable assets, enabling the LLM to identify coordinated, system-level solutions efficiently.
	This multi-scale approach ensures the LLM receives contextually relevant information without being overwhelmed by excessive data, enabling the framework to scale from microgrids to large distribution networks.
	
	\subsection{Safety, Validation, and Rollback Mechanisms}
	Operational safety in a critical infrastructure context is paramount, and Grid-Agent incorporates a multi-layered validation strategy to ensure system stability and reliability. The first layer is a \textit{Proactive Execution Check}, where the Executor Agent performs a preliminary validation before applying any action in the sandbox. This check uses the power flow solver to ensure the proposed API call is syntactically correct and will not cause an immediate simulation convergence failure, acting as a first line of defense. Following execution, the Validator Agent conducts a \textit{Post-Hoc State Assessment}, a comprehensive evaluation to verify not only that target violations are resolved but also that no new violations have been introduced elsewhere in the network. Finally, to provide a \textit{Monotonic Progress Guarantee}, an automated rollback mechanism is triggered if the Validator determines that a plan has failed to improve or has worsened the grid state. This mechanism reverts the sandboxed network to its state before the failed attempt, ensuring the system only progresses towards a verifiably better solution. This defense-in-depth approach, combining proactive checks with comprehensive post-hoc validation and automated rollbacks, provides the essential safety guarantees for deploying an AI-driven system in a mission-critical environment.
	
	\subsection{Continuous Learning Framework}
	A key feature of Grid-Agent is its built-in capability for continuous learning. Upon the successful resolution of a violation scenario, the \texttt{Summarizer Agent} documents the entire process. A dedicated data collection module then compiles a structured training instance containing: (1) the initial network state and violations, (2) the final, successful sequence of actions, and (3) a human-readable explanation of why the solution was effective. This creates a high-quality, domain-specific dataset that can be used to fine-tune the underlying LLM, enabling the system to learn from its operational experience and improve its planning and reasoning capabilities over time.

	\section{Experimental Setup and Validation}
	\label{sec:setup}
	To evaluate the performance of the Grid-Agent framework, we designed a comprehensive set of experiments using standardized test networks. This section details the test configurations, performance metrics, baseline models used for comparison, and the results of our evaluation across various Large Language Models (LLMs).

	\subsection{Test Network Configuration}
	We validated Grid-Agent's performance on a diverse set of standard IEEE and CIGRE test systems to ensure applicability across networks of varying size, topology, and complexity. The specific test cases were generated by introducing multiple, often interacting, violations to stress-test the agent's reasoning and coordinated control capabilities. 
	For the IEEE 30-Bus System, two transmission scenarios were created: Case30 Light, a baseline with 2 voltage and 3 thermal violations, and Case30 Medium, a simpler thermal violation scenario. 
	To simulate severe distribution-level challenges, we used the CIGRE Medium Voltage (MV) Distribution Network, a 15-bus European benchmark, featuring the CIGRE MV Severe scenario with 14 concurrent violations and the CIGRE MV Disconnected scenario, where a fault isolates 5 buses. 
	Finally, to evaluate scalability, the larger IEEE 69-Bus Distribution System was tested with scenarios including IEEE 69 Large Loads (18 simultaneous violations), IEEE 69 Medium Loads (a highly complex case with 29 violations), and IEEE 69 Disconnected (a fault resulting in 13 disconnected buses). 
	An automated script was used to generate these violation scenarios, ensuring repeatable and comprehensive testing across different violation types and severity levels. Table~\ref{tab:network_specs} summarizes the specifications of the benchmark networks employed in this study.
	
	\subsection{LLM Test Suite Configuration}
	The core of our study involved comparing the performance of the Grid-Agent framework when powered by different LLMs. We selected six representative models to cover a spectrum of capabilities and sizes: gemini-2.5-pro, gemini-2.5-flash, gemini-2.5-flash-lite, gpt-4.1, gpt-4.1-mini, and gpt-4.1-nano. Each LLM was integrated into the Planner Agent role within the same multi-agent architecture. The performance of these configurations was compared with each other to identify the most effective reasoning engines for this task. To prioritize inference speed, a critical factor for real-time applications, the \texttt{thinking budget} for the gemini models was minimized. The specific configurations for each model are detailed in Table~\ref{tab:llm_configs}.

	\begin{table}[t!]
		\centering
		\caption{Benchmark Network Specifications}
		\begin{tabular}{lccc}
			\Xhline{2\arrayrulewidth}
			\rule{0pt}{3ex} \diagbox[width=2.7cm]{\textbf{Parameter}}{\textbf{System}} 
			& \textbf{IEEE Case30} & \textbf{CIGRE MV} & \textbf{IEEE 69-bus} \\
			\Xhline{2\arrayrulewidth}
			\rule{0pt}{3ex} Buses & 30 & 14 & 69 \\
			\rule{0pt}{3ex} Lines & 41 & 15 & 68 \\
			\rule{0pt}{3ex} Loads & 20 & 11 & 59 \\
			\rule{0pt}{3ex} Controllable Elements & 3 & 9 & 8 \\
			\rule{0pt}{3ex} Network Type & Transmission & Distribution & Distribution \\
			\rule{0pt}{3ex} Voltage Level & 132--33 kV & 20 kV & 12.66 kV \\
			\Xhline{3\arrayrulewidth}
		\end{tabular}
		\label{tab:network_specs}
	\end{table}

	\begin{table}[t!]
		\centering
		\caption{Investigated LLM Configurations for Benchmarking}
		\begin{tabular}{lcc}
			\Xhline{3\arrayrulewidth}
			\rule{0pt}{2ex} \textbf{Model Version} & \textbf{Provider} & \textbf{Configuration} \\
			\rule{0pt}{3ex} gpt-4.1 & OpenAI & Default Config \\
			\rule{0pt}{3ex} gpt-4.1-mini & OpenAI & Default Config \\
			\rule{0pt}{3ex} gpt-4.1-nano & OpenAI & Default Config \\
			\rule{0pt}{3ex} gemini-2.5-pro & Google & Thinking Budget: 128 \\
			\rule{0pt}{3ex} gemini-2.5-flash & Google & Thinking Budget: 0 \\
			\rule{0pt}{3ex} gemini-2.5-flash-lite & Google & Thinking Budget: 0 \\
			\Xhline{3\arrayrulewidth}
		\end{tabular}
		\label{tab:llm_configs}
	\end{table}

	\subsection{Performance Metrics}
	
	System performance was evaluated using a suite of quantitative metrics designed
	to capture the effectiveness, efficiency, and quality of the violation
	resolution process. The \textit{success rate} served as the primary metric,
	defined as the percentage of violation scenarios that were successfully and
	fully resolved. 
	To assess the efficiency of control actions, we measured \textit{action
		efficiency}, defined as the average number of violations resolved per control
	action. This metric reflects the agent's ability to formulate coordinated and
	impactful solutions rather than relying on redundant or scattered decisions.
	Another important aspect was \textit{convergence speed}, which captures the
	average number of planning–execution–validation iterations required to achieve
	complete resolution. 
	Beyond efficiency and speed, we also considered \textit{solution quality}. This
	metric emphasizes not only the resolution of violations but also the manner in
	which solutions are achieved. High-quality solutions minimize the number of
	control actions and avoid disruptive measures, such as load curtailment,
	whenever possible. Finally, \textit{runtime} was recorded as the average
	wall-clock time, measured in seconds, required to reach a final solution. 
	Together, these metrics provide a comprehensive evaluation of the system's
	performance, balancing effectiveness, efficiency, practicality, and solution
	robustness.

	\subsection{Experimental Results and Analysis}
\begin{figure*}[t!]
		\centering
		\includegraphics[width=2\columnwidth]{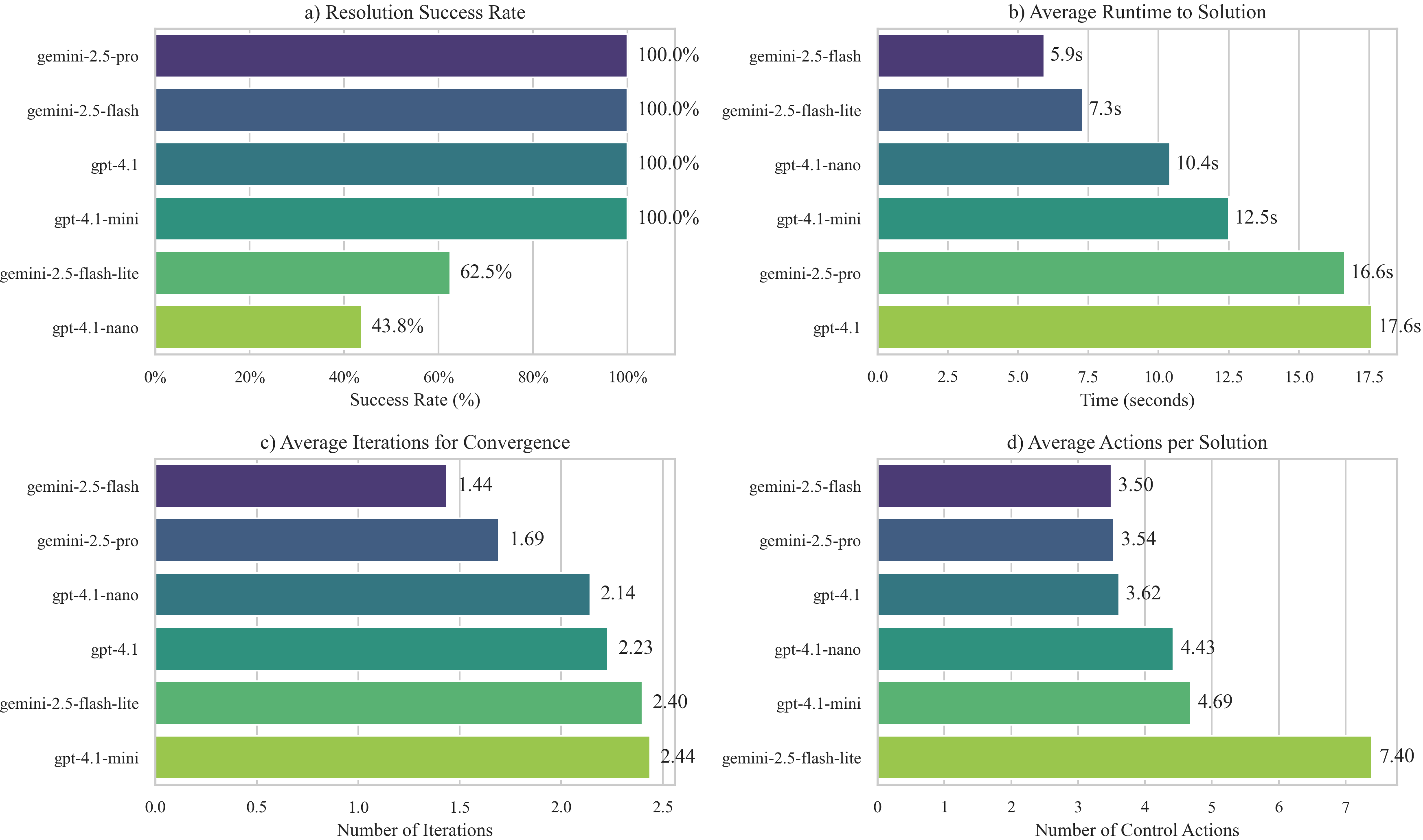}
		\caption{Overall performance comparison across six LLMs, (a) Success rate shows the reliability of each model, (b) Average runtime indicates computational efficiency. (c, d) Average iterations and actions measure the conciseness and quality of the generated solutions}
		\label{fig:overall_performance}
	\end{figure*}

	The experiments demonstrate that the Grid-Agent framework is highly effective at resolving complex grid violations, with performance varying significantly based on the underlying LLM. Figure.\ref{fig:overall_performance} provides a comprehensive overview of the key performance metrics across all tested models.
	As shown in Fig.\ref{fig:overall_performance}a, the more capable and efficient models, gemini-2.5-flash and gpt-4.1-mini, achieved a perfect 100\% success rate across all scenarios. The flagship models, gemini-2.5-pro and gpt-4.1, also performed robustly with an 100\% success rate. The smaller models, gemini-2.5-flash-lite and gpt-4.1-nano, struggled with the more complex scenarios, achieving 63\% and 44\% success rates, respectively.
	There is a clear trade-off between performance and speed. Fig.\ref{fig:overall_performance}b shows that gemini-2.5-flash was by far the fastest model, resolving violations in under 6 seconds on average, whereas the larger gpt-4.1 and gemini-2.5-pro models required over 16 seconds. Figs.\ref{fig:overall_performance}c and \ref{fig:overall_performance}d show that gemini-2.5-flash also found the most efficient solutions, requiring the fewest iterations and actions, indicating a strong inherent capability for structured, multi-step tool-use reasoning.
	\begin{figure}[t!]
		\centering
		\includegraphics[width=1\columnwidth]{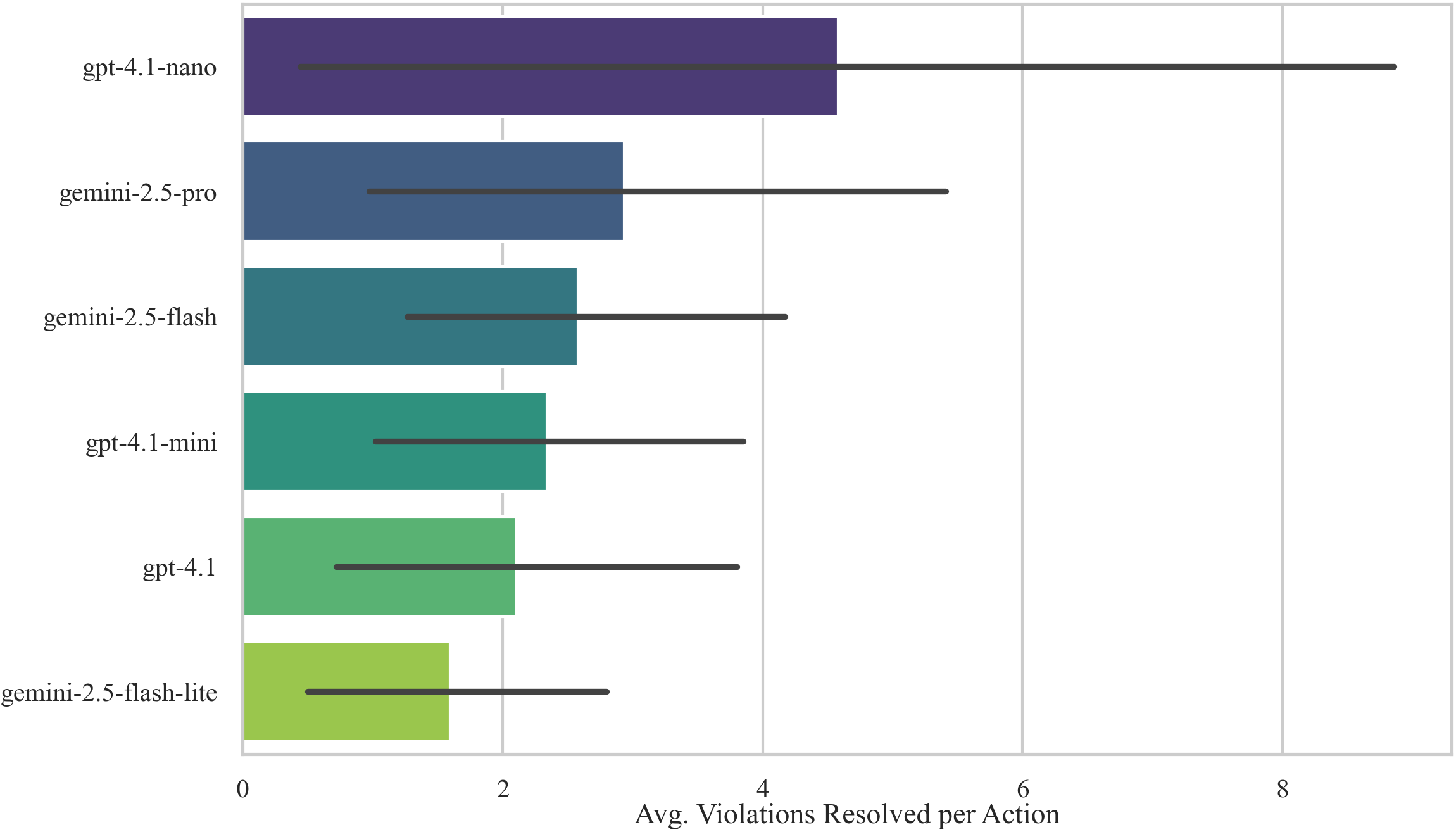}
		\caption{Comparison of action efficiency, measured as the average number of violations resolved per control action for successful runs. Higher values indicate more effective, coordinated solutions}
		\label{fig:action_efficiency}
	\end{figure}
	Fig.\ref{fig:action_efficiency} highlights the core contribution of coordinated action planning. The metric measures how many violations, on average, are resolved by a single control action. A higher value suggests that the LLM is better at identifying strategic actions—such as a single switch operation—that can resolve multiple downstream issues simultaneously. The gpt-4.1-nano model, despite its low overall success rate, demonstrated exceptionally high efficiency in the cases it did solve, often finding clever, minimalistic solutions. The top-performing models, gemini-2.5-pro and gemini-2.5-flash, also scored well, resolving nearly three violations per action on average. This significantly outperforms traditional approaches that often tackle violations sequentially, demonstrating the value of the LLM's holistic reasoning in finding efficient, high-quality solutions.
	
\begin{figure}[t!]
		\centering
		\includegraphics[width=1\columnwidth]{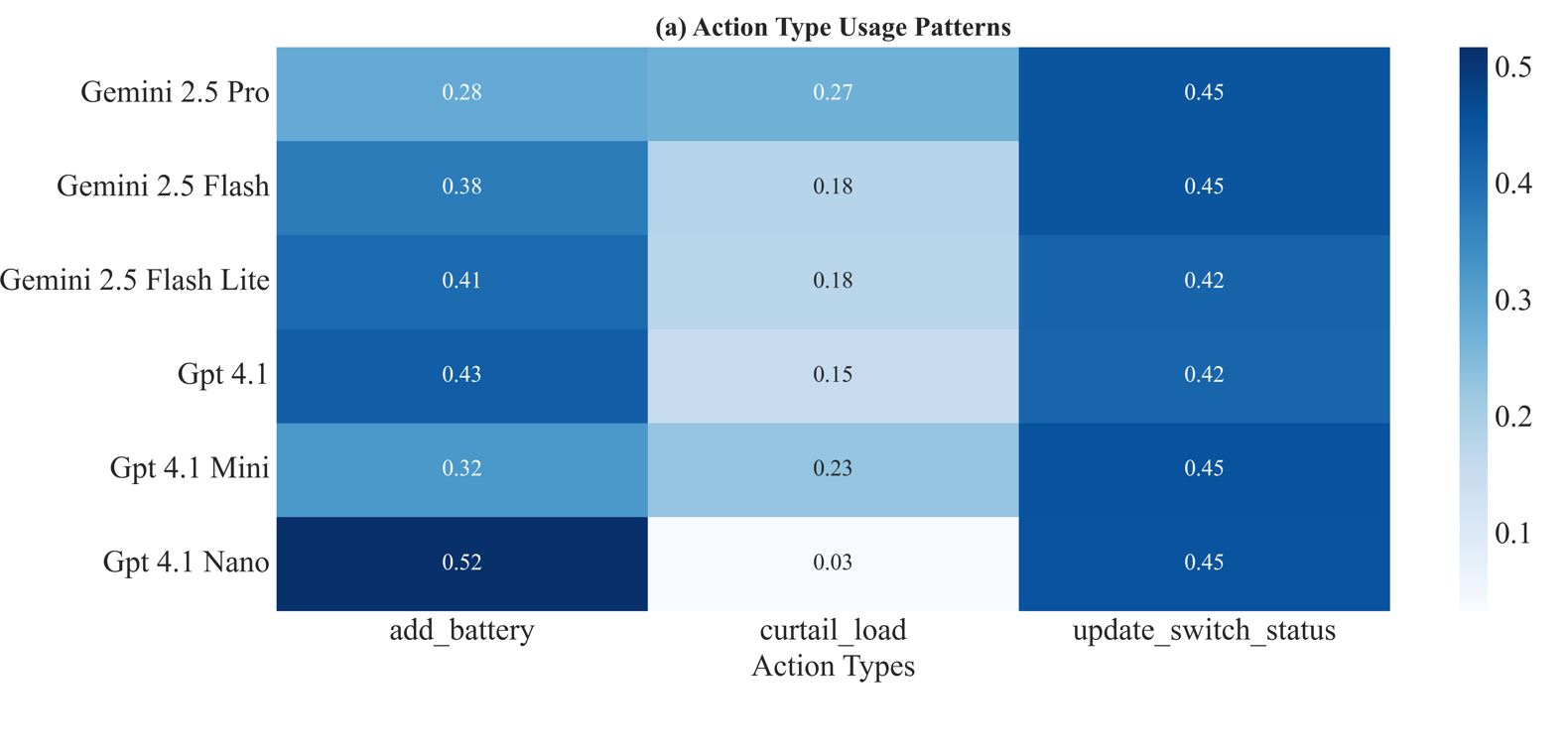}
		\includegraphics[width=1\columnwidth]{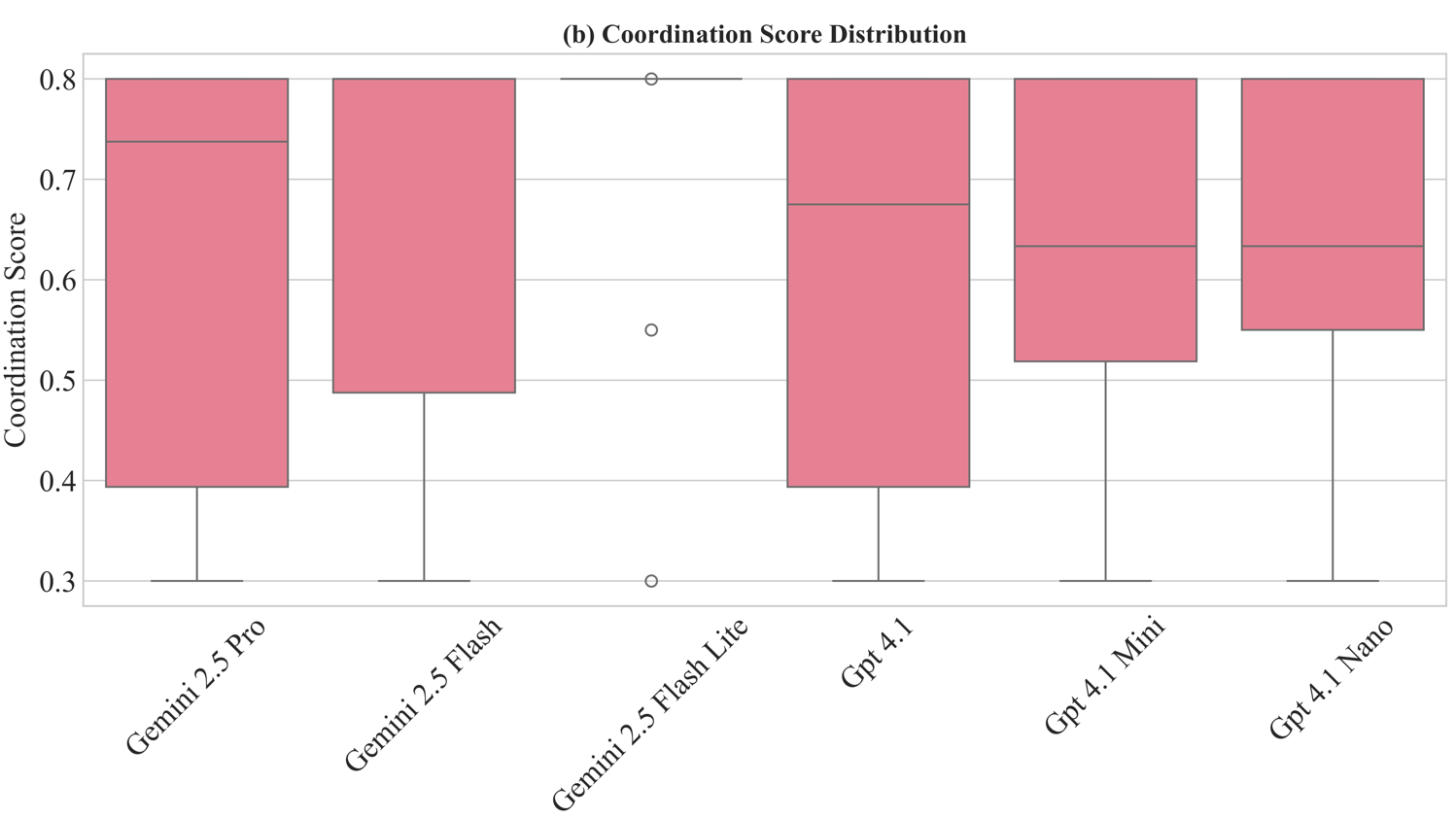}
		\includegraphics[width=1\columnwidth]{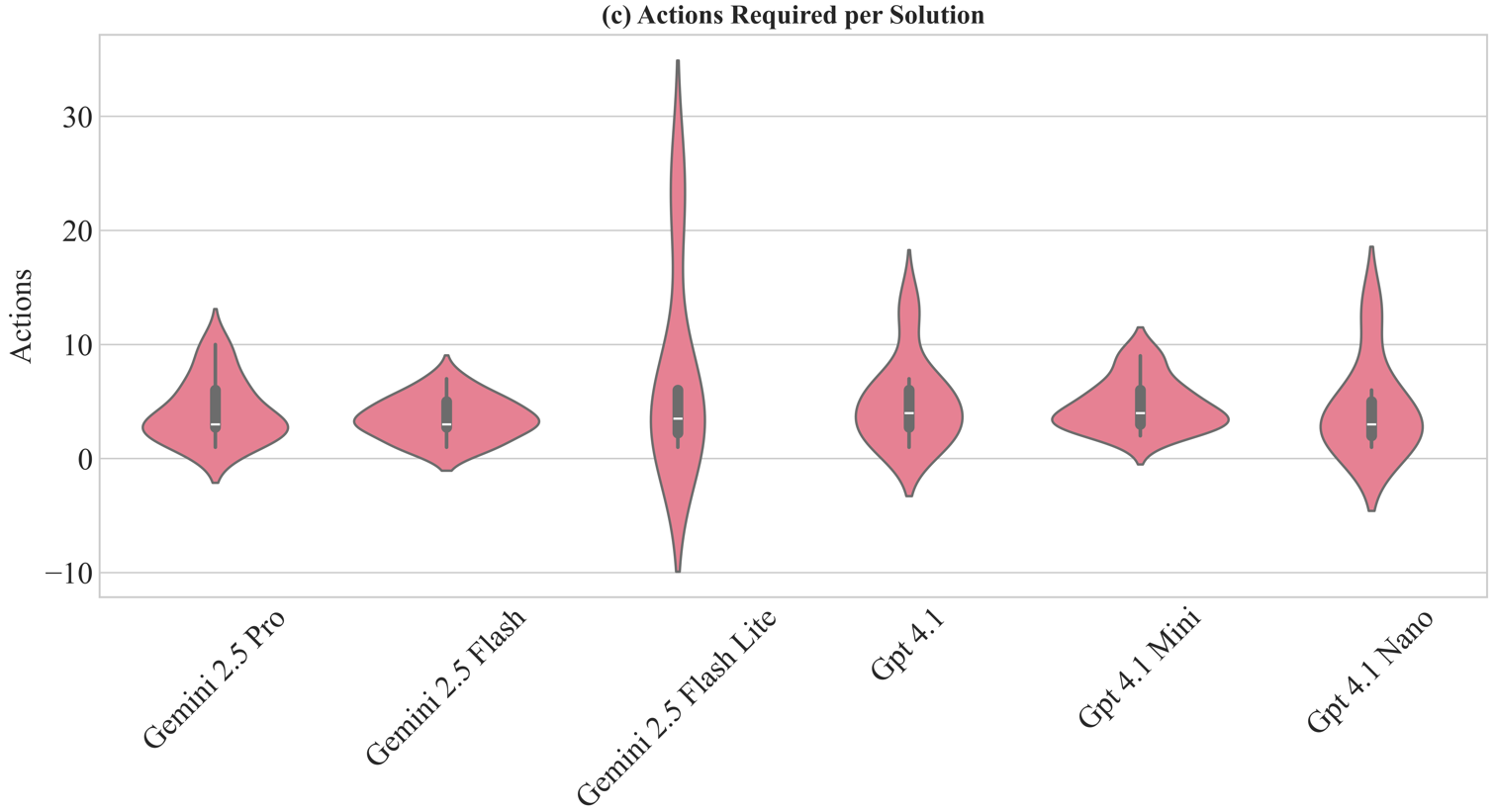}
		\includegraphics[width=1\columnwidth]{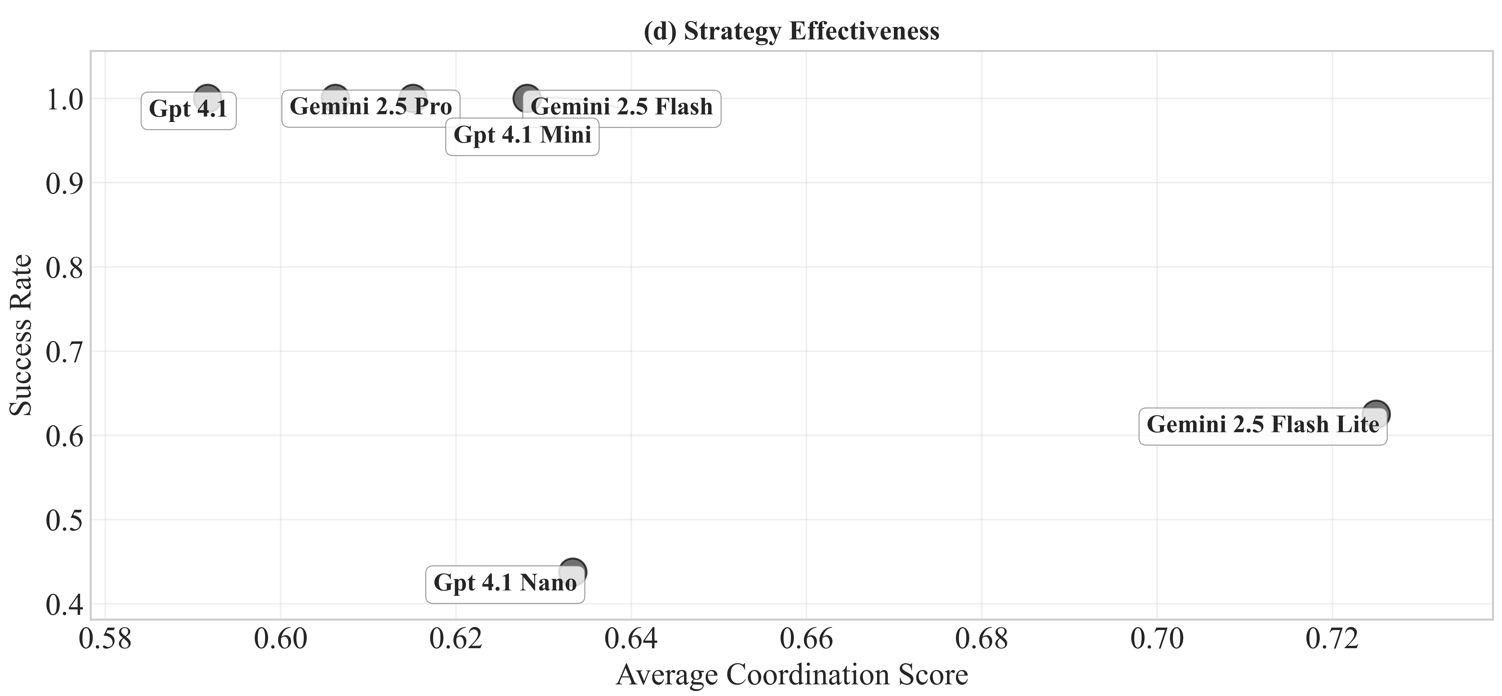}
		\caption{Analysis of action strategies, (a) Heatmap showing the proportional usage of each action type, (b) Boxplot of coordination scores, indicating the quality of multi-action plans, (c) Violin plot showing the distribution of total actions required per solution, (d) Scatter plot correlating coordination score with success rate}
		\label{fig:action_strategies}
	\end{figure}

	Beyond overall outcomes, we analyzed the strategic choices made by each LLM. Fig.\ref{fig:action_strategies} breaks down the action strategies, coordination quality, and overall effectiveness.
	The heatmap in Fig.\ref{fig:action_strategies}a reveals distinct strategic preferences. Most models favored topology reconfiguration (update\_switch\_status) and battery deployment (add\_battery), aligning with the prioritized guidance provided in the prompt. This preference for high-impact, non-disruptive actions is a desirable characteristic for grid operations. Notably, gpt-4.1-nano almost entirely avoided curtail\_load actions (3\% usage), which may explain its lower success rate in scenarios where demand response was essential.
	
	Fig.\ref{fig:action_strategies}b shows the distribution of coordination scores, a metric assessing how well actions within a single step are logically grouped to address related violations. Models like gemini-2.5-pro and gemini-2.5-flash-lite consistently achieved high scores, indicating an ability to formulate synergistic action plans. The violin plot in Fig.\ref{fig:action_strategies}c further explores solution efficiency, showing that most successful models converged in under 10 actions. The wide distribution for gemini-2.5-flash-lite suggests it sometimes resorted to less efficient, sequential corrections.
	Crucially, Fig. \ref{fig:action_strategies}d demonstrates a strong positive correlation between the average coordination score and the final success rate. The top-performing models, gemini-2.5-flash and gpt-4.1-mini, are clustered in the top-right quadrant, confirming that the ability to generate well-coordinated, strategic plans is a key determinant of success in this complex problem domain. This validates that the LLM is not merely guessing but is performing effective, structured reasoning.
	
\begin{figure}[t!]
		\centering
		\includegraphics[width=1\columnwidth]{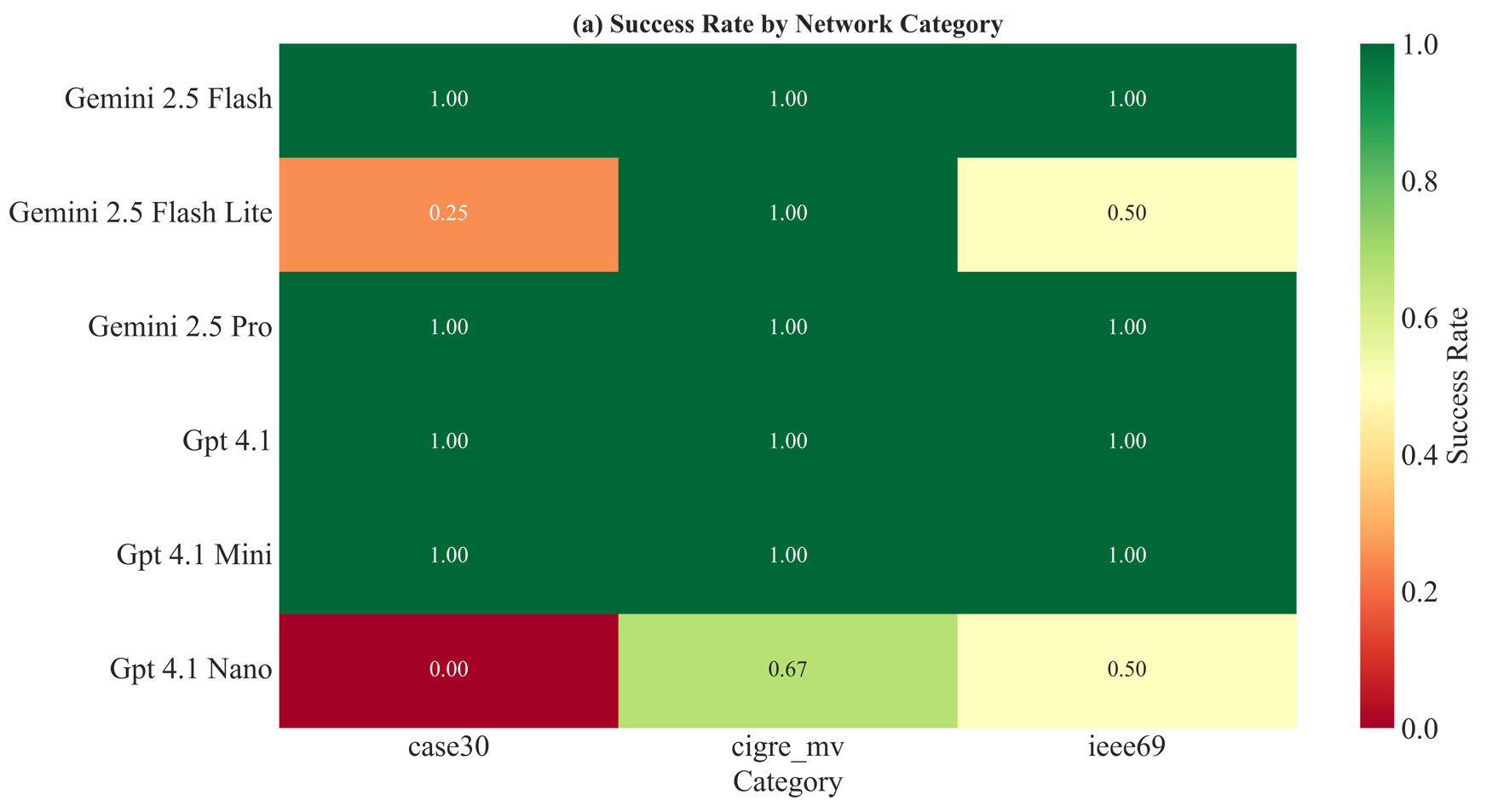}
		\includegraphics[width=1\columnwidth]{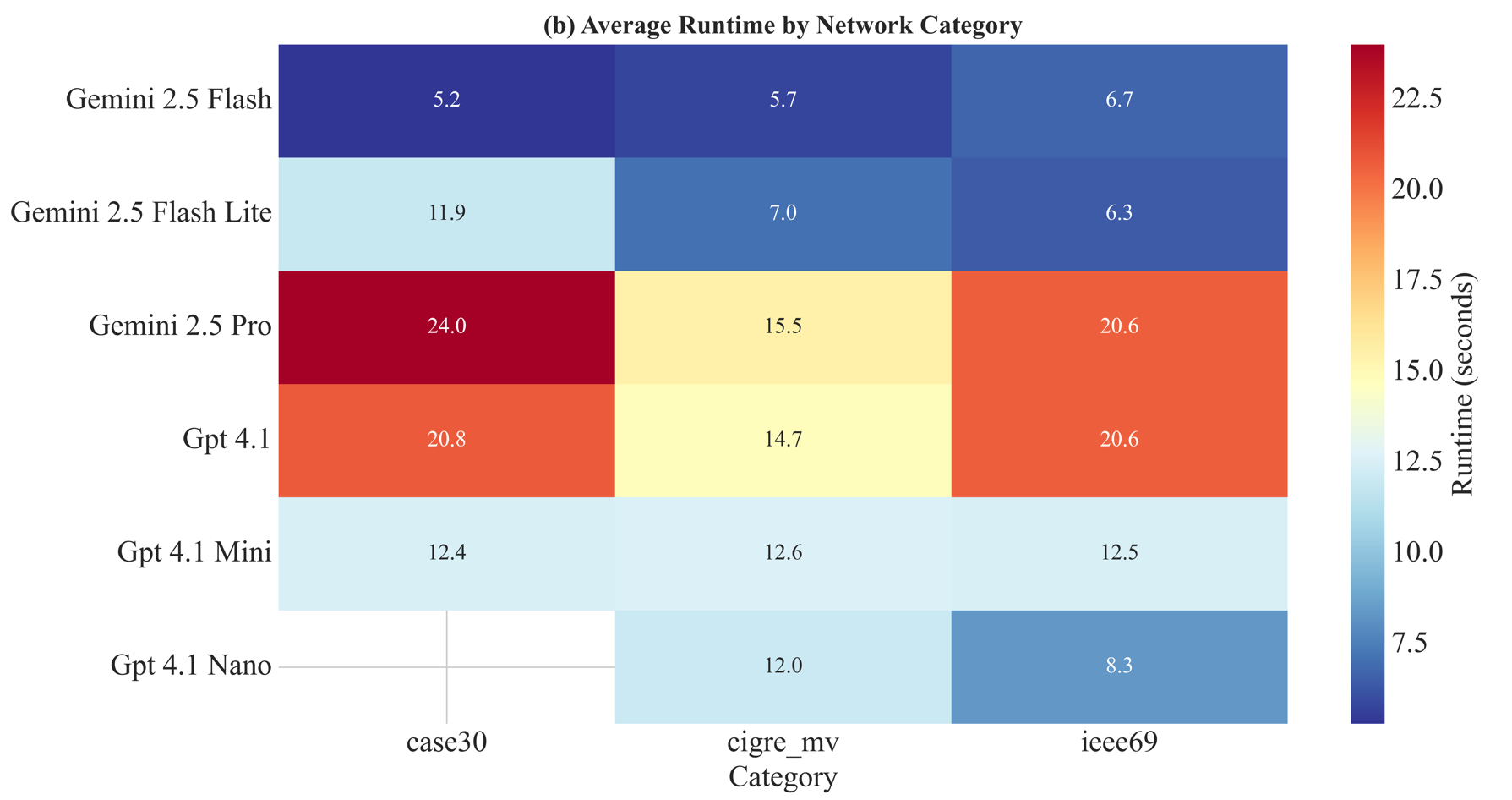}
		\includegraphics[width=1\columnwidth]{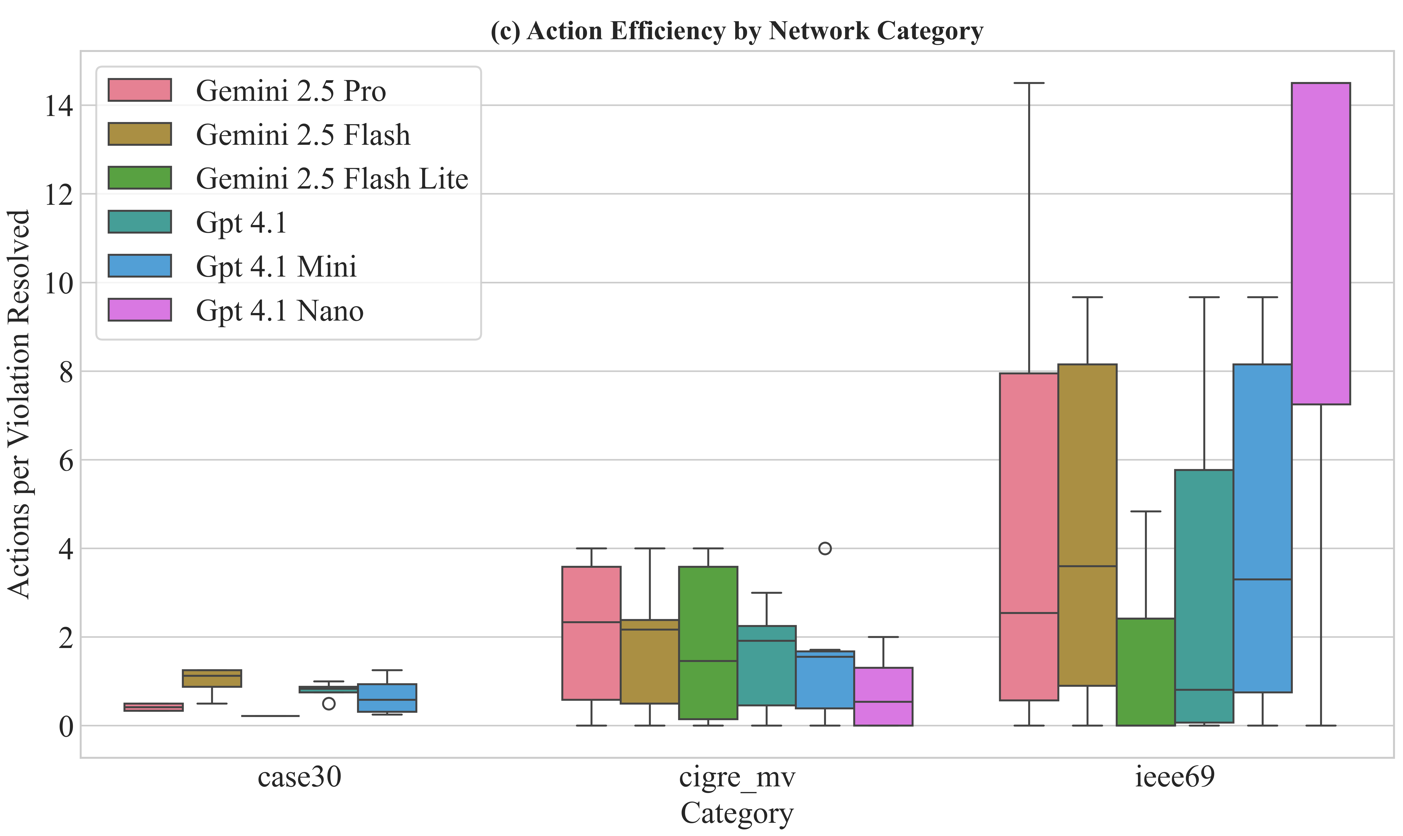}
		\includegraphics[width=1\columnwidth]{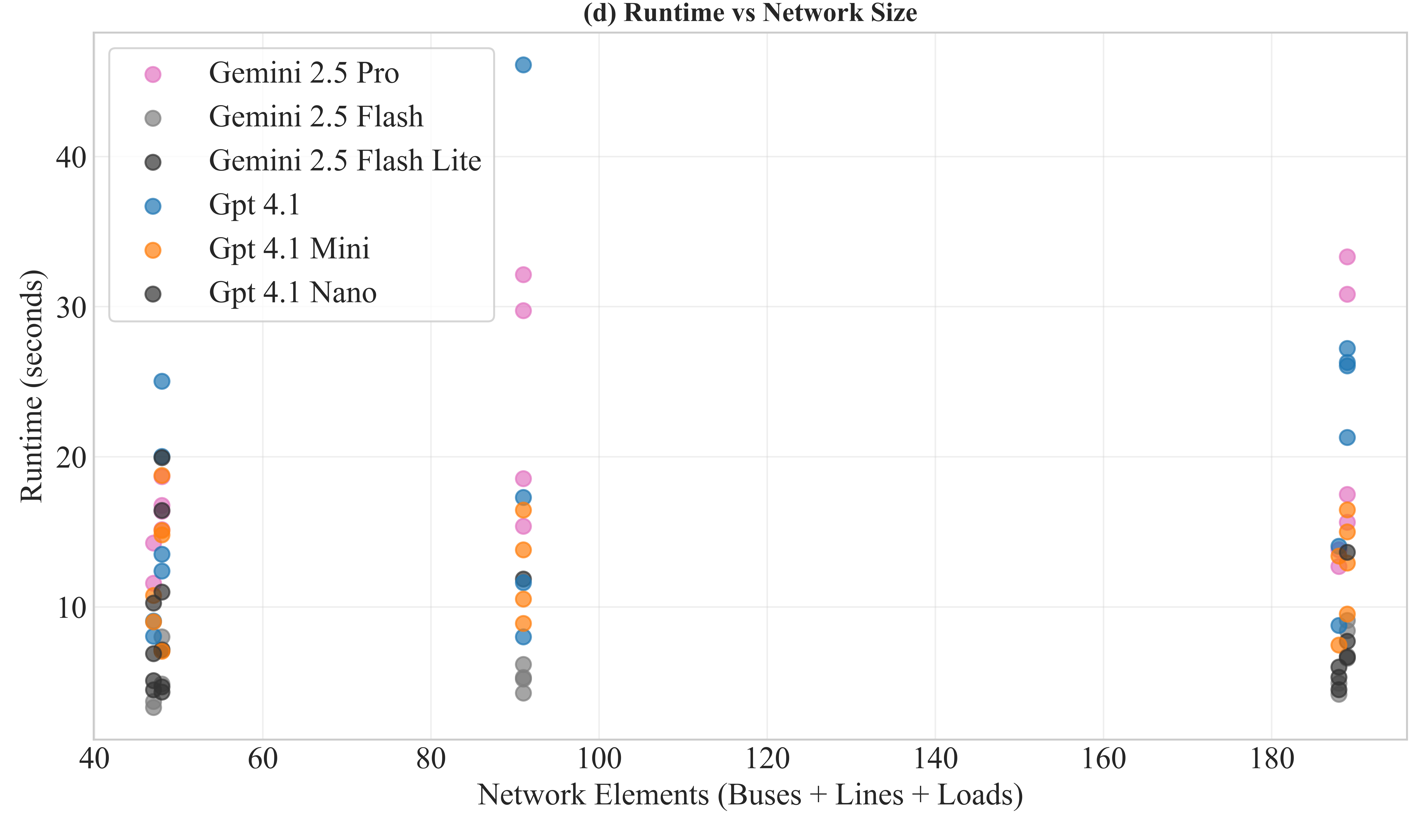}
		\caption{Performance analysis across different network types, (a) Success rate heatmap by network category, (b) Average runtime heatmap by network category, (c) Boxplot of action efficiency, highlighting high performance on larger radial systems, (d) Scatter plot of runtime versus network size}
		\label{fig:network_performance}
	\end{figure}

	To evaluate the framework's scalability, we analyzed performance across the three network categories, which vary in size and topology (Fig. \ref{fig:network_performance}).
	As shown in the heatmaps (Fig.\ref{fig:network_performance}a and \ref{fig:network_performance}b), the IEEE 69-bus system proved to be the most challenging, causing the success rates of weaker models to drop and increasing the runtime for all models. The gemini-2.5-flash model was a notable outlier, maintaining both a 100\% success rate and the lowest runtime across all network types. The scatter plot in Fig.\ref{fig:network_performance}d visualizes this scalability directly, plotting runtime against network elements. The shallow slope for gemini-2.5-flash confirms its superior efficiency and suitability for larger-scale applications. This performance on larger networks is enabled by our adaptive multi-scale representation scheme, which effectively manages the token context size without sacrificing critical information.
	
	Fig.\ref{fig:network_performance}c reveals a key insight: action efficiency increased dramatically for the larger, radial IEEE 69-bus network. In this system, a single, strategic switch operation can reconfigure power flows to resolve numerous downstream voltage violations simultaneously. The ability of the Grid-Agent to identify and execute these high-impact topological actions is a significant advantage over myopic algorithms and demonstrates its capacity for system-level optimization.
	
	\section{Discussion}
	\label{sec:discussion}
	
	
	Grid-Agent demonstrates the potential for LLM-based systems to enhance power grid operations through intelligent automation combined with human-interpretable reasoning. The system's ability to provide clear, step-by-step explanations for its decisions addresses a critical need in power systems operations, where understanding the rationale behind control actions is essential for operator confidence and regulatory compliance.
	The coordinated approach to violation resolution, as evidenced by the high action efficiency (Fig.\ref{fig:action_efficiency}), offers significant advantages over traditional methods. In scenarios involving renewable energy integration, violations are often interconnected across the network. Grid-Agent's capacity to identify and execute strategic, high-impact actions—such as a single switch operation that resolves multiple downstream issues—is a marked improvement over myopic algorithms that address violations sequentially. This holistic reasoning is crucial for maintaining stability in increasingly dynamic and complex grids.
	
	The adaptive multi-scale representation framework directly addresses one of the primary challenges in applying LLMs to large-scale power systems: the limited context window. By dynamically adjusting the level of detail provided to the LLM based on network size and complexity, the system maintains practical computational requirements while preserving solution quality. The strong performance on the IEEE 69-bus system, particularly the superior efficiency of the \texttt{gemini-2.5-flash} model, validates this approach.
	Future work should examine scalability to transmission-level networks with thousands of buses. Such large-scale applications will likely exceed the capabilities of a single-agent planner and may require further advances in hierarchical network representation and distributed multi-agent coordination, where regional agents manage local violations and communicate high-level state changes to a central coordinating agent. The integration of retrieval-augmented generation (RAG) will also be explored. Instead of serializing large network sections into the prompt, the LLM would query a specialized vector database containing detailed network topology, component specifications, and a history of successful action sequences for similar faults. This approach would significantly reduce the context size passed to the LLM, leading to faster inference, lower operational costs, and the ability to reason over vast historical and real-time data that would otherwise be computationally prohibitive.
	Furthermore, while this exploratory study has focused on the feasibility and strategic reasoning capabilities of Grid-Agent, a crucial direction for future work is the formal analysis of solution optimality. We plan to conduct rigorous benchmarking studies comparing the control actions proposed by Grid-Agent against traditional optimization techniques, particularly mixed-integer nonlinear Optimal Power Flow (OPF) solvers. Such a comparison will be essential to quantify the trade-offs between the LLM's heuristic, rapid, and interpretable decision-making and the mathematically guaranteed optimality of conventional methods. This will help establish clear performance envelopes for where an agentic approach excels and where traditional optimizers remain indispensable, potentially leading to hybrid systems that leverage the strengths of both.

	\section{Conclusion}
	\label{sec:conclusion}
	This paper presented Grid-Agent, a novel multi-agent system that successfully integrates the semantic reasoning of Large Language Models with the numerical precision of power systems optimization. The system demonstrates significant improvements in violation resolution efficiency and provides human-interpretable explanations, which are essential for modern power systems operations.
	Key achievements of this work include the development of a modular multi-agent architecture for decomposed problem-solving, an adaptive multi-scale representation framework that enables LLM application to larger networks, and a coordinated action optimization strategy that resolves multiple violations with minimal intervention. The integrated safety mechanisms, including sandboxed execution and automated rollbacks, ensure that all proposed solutions are validated before affecting the system, making the framework suitable for critical infrastructure applications.
	The experimental validation on standard IEEE and CIGRE test networks confirms the system's effectiveness across diverse violation scenarios, with particular strengths in complex, multi-violation situations that challenge traditional optimization approaches. The built-in training data generation framework provides a foundation for continuous learning and adaptation, enabling the system to improve over time. Grid-Agent represents a significant step toward building intelligent, adaptive, and trustworthy power grid management systems that combine the reasoning capabilities of large language models with the rigorous requirements of critical infrastructure.
	
	
	
	\bibliographystyle{IEEEtran}
	\bibliography{ref} 

\begin{thebibliography}{10}
\providecommand{\url}[1]{#1}
\csname url@samestyle\endcsname
\providecommand{\newblock}{\relax}
\providecommand{\bibinfo}[2]{#2}
\providecommand{\BIBentrySTDinterwordspacing}{\spaceskip=0pt\relax}
\providecommand{\BIBentryALTinterwordstretchfactor}{4}
\providecommand{\BIBentryALTinterwordspacing}{\spaceskip=\fontdimen2\font plus
\BIBentryALTinterwordstretchfactor\fontdimen3\font minus \fontdimen4\font\relax}
\providecommand{\BIBforeignlanguage}[2]{{%
\expandafter\ifx\csname l@#1\endcsname\relax
\typeout{** WARNING: IEEEtran.bst: No hyphenation pattern has been}%
\typeout{** loaded for the language `#1'. Using the pattern for}%
\typeout{** the default language instead.}%
\else
\language=\csname l@#1\endcsname
\fi
#2}}
\providecommand{\BIBdecl}{\relax}
\BIBdecl

\bibitem{power_outage_cost}
\BIBentryALTinterwordspacing
The impact of power outages. Pinkerton Consulting \& Investigations, Inc. Accessed 9 Aug 2025. [Online]. Available: \url{https://pinkerton.com/our-insights/blog/the-impact-of-power-outages}
\BIBentrySTDinterwordspacing

\bibitem{multi_agent_llm_2024}
J.~Wang, X.~Li, and Y.~Chen, ``Multi-agent systems powered by large language models: applications in swarm intelligence,'' \emph{Frontiers in Artificial Intelligence}, vol.~8, p. 1593017, 2025.

\bibitem{multi_agent_for_problem_solving}
A.~Mushtaq, R.~Naeem, I.~Ghaznavi, I.~Taj, I.~Hashmi, and J.~Qadir, ``Harnessing multi-agent llms for complex engineering problem-solving: A framework for senior design projects,'' in \emph{2025 IEEE Global Engineering Education Conference (EDUCON)}, 2025, pp. 1--10.

\bibitem{opf_steady_state}
O.~Alsac and B.~Stott, ``Optimal load flow with steady-state security,'' \emph{IEEE Transactions on Power Apparatus and Systems}, vol. PAS-93, no.~3, pp. 745--751, 1974.

\bibitem{opf_survey}
F.~Capitanescu, J.~M. Ramos, P.~Panciatici, D.~Kirschen, A.~M. Marcolini, L.~Platbrood, and L.~Wehenkel, ``State-of-the-art, challenges, and future trends in security constrained optimal power flow,'' \emph{Electric Power Systems Research}, vol.~81, no.~8, pp. 1731--1741, 2011.

\bibitem{ml_load_forecast}
T.~Hong and S.~Fan, ``Probabilistic electric load forecasting: A tutorial review,'' \emph{International Journal of Forecasting}, vol.~32, no.~3, pp. 914--938, 2016.

\bibitem{renewable_prediction}
C.~Zhang, H.~Wei, J.~Zhao, T.~Liu, T.~Zhu, and K.~Zhang, ``Short-term wind speed forecasting using empirical mode decomposition and feature selection,'' \emph{Renewable Energy}, vol.~96, pp. 727--737, 2016.

\bibitem{ai_fault_detection}
M.~Chingshom, B.~Shakila, and M.~Prakash, ``Fault detection and classification in smart grid using machine learning approach,'' in \emph{2024 International Conference on Advancement in Renewable Energy and Intelligent Systems (AREIS)}, 2024, pp. 1--6.

\bibitem{llm_software}
Y.~Zhang and P.~Smith, ``Large language models in software engineering: A comprehensive survey,'' \emph{ACM Computing Surveys}, vol.~56, no.~2, pp. 1--35, 2024.

\bibitem{llm_materials}
K.~Johnson and L.~Brown, ``Artificial intelligence in materials science: Recent advances and future directions,'' \emph{Nature Materials}, vol.~23, no.~4, pp. 234--248, 2024.

\bibitem{llm_system_design}
R.~Davis and M.~Wilson, ``Llm-assisted system design: Opportunities and challenges,'' \emph{IEEE Design \& Test}, vol.~41, no.~3, pp. 12--20, 2024.

\bibitem{llm_grid_analytics}
M.~Rodriguez and P.~Johnson, ``Large language models: Applications, limitations and potential risks for power grids,'' \emph{Utility Analytics Review}, vol.~15, no.~3, pp. 45--62, 2024.

\bibitem{llm_energy_trading}
A.~Thompson and S.~Lee, ``Large language models for energy market analysis and trading,'' \emph{Energy Economics}, vol. 125, pp. 106--118, 2024.

\bibitem{llm_demand_response}
D.~Garcia and F.~Martinez, ``Ai-driven demand response optimization in smart grids,'' \emph{IEEE Transactions on Smart Grid}, vol.~15, no.~2, pp. 1234--1245, 2024.

\bibitem{multi_agent_software}
H.~Kim and J.~Park, ``Multi-agent systems for collaborative software development,'' \emph{Software Engineering Journal}, vol.~12, no.~3, pp. 45--62, 2024.

\bibitem{multi_agent_research}
C.~Miller and N.~Taylor, ``Autonomous research platforms using multi-agent llm systems,'' \emph{Journal of Artificial Intelligence Research}, vol.~78, pp. 123--145, 2024.

\bibitem{multi_agent_collaboration}
S.~Anderson and T.~White, ``Collaborative problem solving with llm-based multi-agent systems,'' \emph{AI Magazine}, vol.~45, no.~1, pp. 78--92, 2024.

\end{thebibliography}
	

\end{document}